\documentclass[prl,twocolumn,superscriptaddress,nofootinbib]{revtex4}

\usepackage{color}
\usepackage{graphicx,bm,times}
\usepackage{lipsum}
\usepackage{amsmath}
\usepackage{gensymb}
\usepackage{color,soul, mathrsfs}
\usepackage[]{natbib}
\usepackage{mathtools}
\usepackage[pdftex,colorlinks=true,linkcolor=blue,citecolor=blue,urlcolor=black]{hyperref}
\usepackage{amsmath, amsthm, amssymb}
\usepackage[caption=false]{subfig}
\usepackage{comment}
\usepackage{url}
\usepackage{tikz}
\usepackage{bbold}
\usepackage{balance}
\usepackage{siunitx}
\usepackage{braket}
\usepackage{filecontents}

\DeclareMathOperator{\Var}{Var}
\newcommand{\be}{\begin{equation}}
\newcommand{\ee}{\end{equation}}

\begin{document}

\title{Maximising Precision in Saturation-Limited Absorption Measurements}

\author{Jake Biele}
\email{jb12101@bristol.ac.uk}
\affiliation{Quantum Engineering Technology Labs, H. H. Wills Physics Laboratory and Department of Electrical \& Electronic Engineering, University of Bristol, BS8 1FD, United Kingdom.}
\affiliation{Quantum Engineering Centre for Doctoral Training, H. H. Wills Physics Laboratory and Department of Electrical \& Electronic Engineering, University of Bristol, Tyndall Avenue, BS8 1FD, United Kingdom.}
\author{Sabine Wollmann}
\affiliation{Quantum Engineering Technology Labs, H. H. Wills Physics Laboratory and Department of Electrical \& Electronic Engineering, University of Bristol, BS8 1FD, United Kingdom.}
\author{Joshua W. Silverstone}
\affiliation{Quantum Engineering Technology Labs, H. H. Wills Physics Laboratory and Department of Electrical \& Electronic Engineering, University of Bristol, BS8 1FD, United Kingdom.}
\author{Jonathan C. F. Matthews}
\affiliation{Quantum Engineering Technology Labs, H. H. Wills Physics Laboratory and Department of Electrical \& Electronic Engineering, University of Bristol, BS8 1FD, United Kingdom.}
\author{Euan J. Allen}
\email{euan.allen@bristol.ac.uk}
\affiliation{Quantum Engineering Technology Labs, H. H. Wills Physics Laboratory and Department of Electrical \& Electronic Engineering, University of Bristol, BS8 1FD, United Kingdom.}

\begin{abstract}
Quantum fluctuations in the intensity of an optical probe is noise which limits measurement precision in absorption spectroscopy. Increased probe power can offer greater precision, however, this strategy is often constrained by sample saturation. Here, we analyse measurement precision for a generalised absorption model in which we account for saturation and explore its effect on both classical and quantum probe performance. We present a classical probe-sample optimisation strategy to maximise precision and find that optimal probe powers always fall within the saturation regime. We apply our optimisation strategy to two examples, high-precision Doppler broadened thermometry and an absorption spectroscopy measurement of Chlorophyll A. We derive a limit on the maximum precision gained from using a non-classical probe and find a strategy capable of saturating this bound. We evaluate amplitude-squeezed light as a viable experimental probe state and find it capable of providing precision that reaches to within ${> 85\%}$ of the ultimate quantum limit with currently available technology.
\end{abstract}

\date{\today}

\maketitle

Absorption spectroscopy exploits light-matter interactions to give precise measurements of sample composition. This technique is widely applied with key use-cases including drug analysis~\cite{Evans2014ApplicationsEnvironment}, environmental monitoring~\cite{Burgess1995Absorption-basedSensors}, atomic characterisation~\cite{Yang2007AtomicChip}, and industrial process monitoring~\cite{Scotter1990UseProcesses}. 
In absorption spectroscopy, exceeding sample-specific probe intensities often leads to irreversible damage through a host of saturation-dependent mechanisms~\cite{Neuman1999,Peterman2003}. Investigating probe performance in the saturation regime is therefore crucial for simultaneously optimising performance and minimising irreversible damage~\cite{Carpentier1987,Bopp1997,NathanHenderson2007} of delicate samples such as archaeological finds, living cells, or food products~\cite{Toffolo2018InfraredATR,Taylor2016, Scotter1990UseProcesses}. 
Intensity dependent quantum noise within the probe scales favourably with probe power but fundamentally limits the precision of absorption measurements. Consequently, the saturation intensity of the sample places a bound on the achievable measurement precision.

By conducting an analysis of how saturation affects measurement precision, we are able to present a probe-sample optimisation scheme to help classical measurements obtain the highest precision possible with a classical probe. The precision is quantified utilising the Fisher information~\cite{Escher2011GeneralMetrology} -- a measure of how much information about an unknown parameter we can extract from the system. We find that the optimal probe power is always in the saturation regime ($\geq$ 50$\%$ of the saturation power) which highlights an inherent trade-off between precision and damage that saturation-limited classical schemes must navigate. This motivates the need to find alternative probe states that provide greater precision per photon.

Effective states for parameter estimation are identified as being non-classical states of light: single-photon states~\cite{Whittaker2017,Sabines-Chesterking2017}, multi-photon states~\cite{Jones2009,Birchall2016BeatingPhotons,Crespi2012MeasuringPhotons}, or squeezed states~\cite{Lawrie2019,Aasi2013,Atkinson2020QuantumLight,Casacio2021Quantum-enhancedMicroscopy} capable of enhancing performance under linear loss or phase for a fixed resource level~\cite{Taylor2016,Monras2007OptimalChannels,Losero2018UnbiasedTwin-beams,Allen2020ApproachingResources}. Results thus far have focused on the linear absorption regime, with the exception of work by Mitchell~\cite{Mitchell2017Number-unconstrainedSensing} which models the effect of constrained photon number on the performance of Gaussian states for single-parameter estimation. Mitchell numerically explores the performance of Gaussian states for measuring optical depth under a semi-classical model of saturation. Here, we derive an analytical bound on the performance of both Gaussian and non-Gaussian states under saturation. We assess the ability of coherent states, Fock states, and squeezed states to saturating this bound. Our results show that on a per photon basis the Fock state remains optimal for probing in the saturation regime, giving a deeper understanding of when to consider quantum light sources a worthwhile and viable upgrade to saturation-limited measurements.
Additionally, our new theoretical framework opens the door to further analysis of nonlinear absorption spectroscopy schemes that directly employ saturation to enhance image resolution~\cite{Chong2010,Wang2013}. These schemes often incorporate transmission measurements of weak signals into more complex estimators and can therefore build upon the model presented here for further optimisation.

In this manuscript, we construct a physical model and utilise the Fisher information (FI) and quantum Fisher Information (QFI) formalisms to investigate classical and quantum probe performance. We derive the FI obtained from classical probing and present a sample-probe optimisation strategy which we later apply to two examples: Doppler broadened thermometry (DBT) and direct Chlorophyll absorption spectroscopy. Further, we perform a QFI analysis and derive a bound on the achievable precision of any single-mode quantum state. Our work shows the FI obtained by the Fock state saturates this bound and is thus optimal. For a given target precision, we show the quantum probe brightness to be of an order of magnitude less than the required classical probe brightness, making it desirable for ultra-sensitive samples. We identify amplitude-squeezed states as a viable route towards quantum precision enhancement for saturation-limited sensors. The analytical results presented are based on a semi-classical approximation of sample saturation. Complementary to these results, we complete a fully-quantum numerical analysis of the effect of saturation on the higher order moments of the probe's photon statistics and find the analytical quantum advantage to be a lower bound on the achievable advantage of employing quantum light.

\begin{figure}[!t]
    \centering
    \includegraphics[width=0.7\columnwidth]{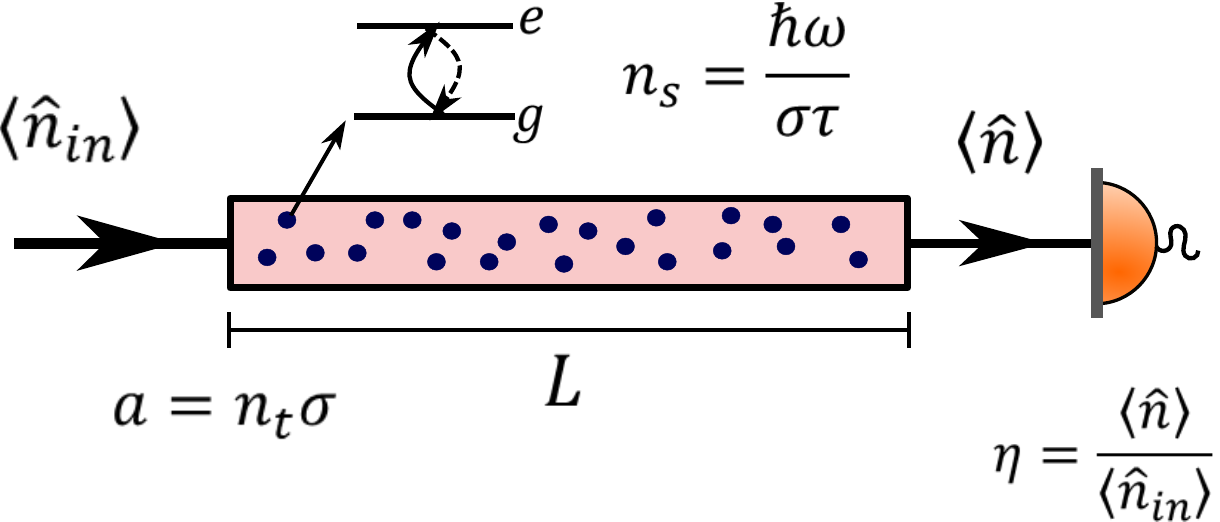}
    \caption{The model schematic: a probe state of mean input photon number ${\langle \hat{n}_{in}\rangle}$ is propagated along a sample of length $L$. The sample consists of homogeneously distributed two-level absorbers with density $n_t$, characteristic transition cross-section $\sigma$, and lifetime $\tau$. The output transmission $\eta$ is measured and used to infer an estimate of the linear absorption coefficient $a$.}
    \label{fig:F1}
\end{figure}

\textit{Fisher Information Formalism.}---The model we consider, depicted in Fig.~\ref{fig:F1}, consists of a saturable absorber of length ${L\in(0,\infty)}$~\si{\centi\meter}, with linear absorption coefficient ${a\in(0,\infty)}$~\si{\centi\meter^{-1}} and saturation intensity $n_s$. The sample is resonantly probed via a single mode with known mean input photon number ${\langle \hat{n}_{in}\rangle}$. Each target absorber is modelled via a static, independent, two-level system. This simplified two-level model can be readily extended to more complex multi-level systems which display a dominant radiative or non-radiative decay path back to ground~\cite{Kobyakov2000,Abitan2008}. The mean photon number is measured at the output to infer a precise estimate of the linear absorption from repeated transmission measurements. Let ${\kappa:=2\langle \hat{n}_{in}\rangle/n_s}$ be the input intensity scaled by the saturation intensity with the saturation regime defined by ${\kappa\geq1}$. Under a steady-state approximation, the sample transmission is given by~\cite{Abitan2008} (see Supplementary Material): \begin{equation}\label{equ:TRANS}
\eta(\kappa, a)=\frac{1}{\kappa}\mathcal{W}[\ln{\kappa}+\kappa-aL].
\end{equation}
Here, ${\mathcal{W}[x]}$ is the Wright Omega function defined over the real line~\cite{Corless2002}. 
We use the standard FI formalism to calculate a bound on the estimate precision of an unknown parameter $x$ encoded in the output state ${\ket{\psi_x}}$. The FI on a target variable $x$ obtained by the set of positive-operator valued measurements (POVMs), ${\hat{\mathbf{M}}=\{m_i\}}$ with ${\sum_i m_i=\mathbb{1}}$, is defined as ${F(x)=\sum_ip(m_i|x)[\partial[\log{p(m_i|x)}]/\partial x]^2}$ \cite{Escher2011GeneralMetrology,Birchall2020QuantumLoss}. The probability measurement outcome $i$ is given by ${p(m_i|x) ={tr\{\ket{\psi_x}\bra{\psi_x}}m_i\}}$. The FI is related to the variance of a given estimator of $x$ via the classical Cram\'{e}r-Rao bound (CRB), inequality~\textit{1} of  Eq.~\ref{equ:CRB}~\cite{Vaart1998AsymptoticStatistics}, which is saturated by an optimal estimator. The quantum Fisher information (QFI) $\mathcal{Q}(x)$ is then defined as the maximum possible FI obtained by optimising over all POVMs and is related to the FI via the quantum Cram\'{e}r-Rao bound (QCRB), inequality~\textit{2} of Eq.~\ref{equ:CRB}~\cite{Braunstein1994StatisticalStates}:
\begin{equation}\label{equ:CRB}
    \frac{1}{\Var(x)}\stackrel{\textit{1}}{\leq} F(x)\stackrel{\textit{2}}{\leq}\mathcal{Q}(x).
\end{equation}

We assume that the saturated loss channel acts like a beam splitter on the optical mode with reflectivity determined by ${1-\eta(\kappa,a)}$. This semi-classical approximation only accounts for the effect of saturation on the first moment of the input state's photon number and thus does not account for the effect on its quantum noise.
\begin{figure*}
    \includegraphics[scale=0.85]{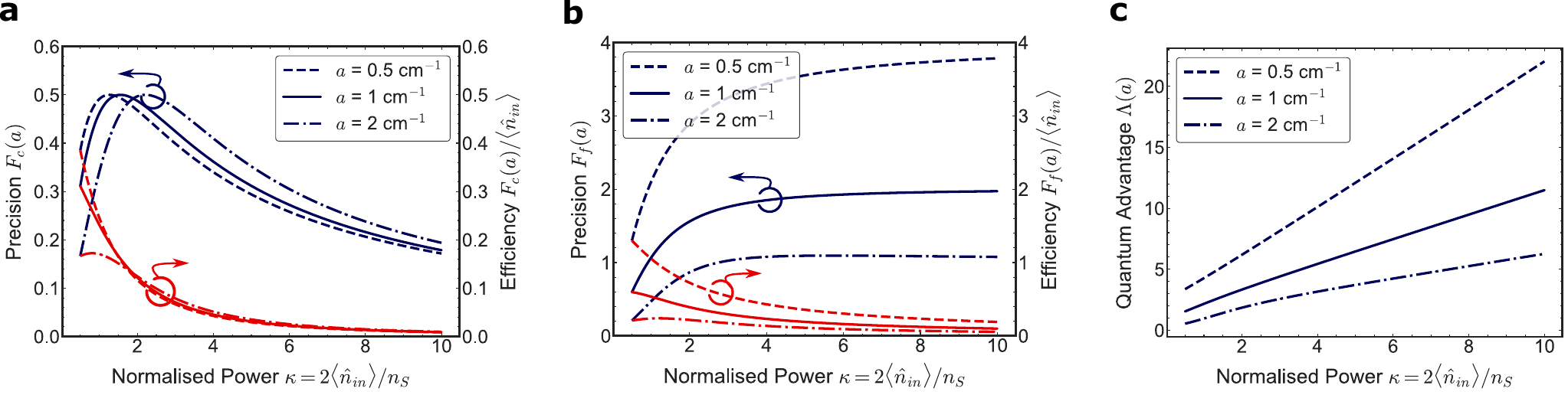}
    \caption{Precision (blue) and efficiency (red) obtained using (a) a classical probe or (b) the optimal quantum probe, calculated across the linear ${(\kappa<1)}$ and saturated ${(\kappa\geq1)}$ pump regimes. (c) shows the quantum advantage $\Lambda$. The sample has a fixed length $L=$~\SI{1}{\centi\meter} and is plotted for absorption coefficients ${a\in\{0.5}$~\si{\centi\meter^{-1}},~$1$~\si{\centi\meter^{-1}},~$2$~\si{\centi\meter^{-1}}$\}$}.
    \label{fig:F2}
\end{figure*}

\textit{Classical probe performance.}---A classical laser probe is well approximated by a coherent state~\cite{Sargent2018LaserPhysics} $\ket{\alpha}$ with ${\langle \hat{n}_{in}\rangle=|\alpha|^2}$ and ${\Var(\hat{n}_{in})=\langle \hat{n}_{in}\rangle}$. In direct absorption schemes, the QCRB is saturated by direct transmission measurements~\cite{Allen2020ApproachingResources}. We calculate ${F(\eta)}$ and relate it to the FI on the linear absorption coefficient, $F(a)$, using the following formula: ${F(a)=(\partial\eta/\partial a)^2 F(\eta)}$. The transmission variance ${\Var(\eta)}$ is related to the output state photon number variance ${\Var(\hat{n})}$ via error propagation~\cite{Ziegel1999TheoryEstimation} (see Supplementary Material B). Combining these two relations gives: \begin{equation}\label{equ:CFI1}
    F(a)=\Big(\frac{\partial\eta}{\partial a}\Big)^2\frac{\langle \hat{n}_{in}\rangle^2}{\Var(\hat{n})}.
\end{equation}
The semi-classical loss approximation allows us to define the output photon number variance analytically via~\cite{Allen2020ApproachingResources}: \begin{equation}\label{equ:LLC}
    \Var(\hat{n}) = \eta^2\Var(\hat{n}_{in}) + \eta(1- \eta)\langle \hat{n}_{in}\rangle.
\end{equation}
For a coherent input state, the FI, ${F_c(a)}$, is given by: \begin{equation}\label{equ:CFI}
     F_c(a)=\Big(\frac{L\eta}{1+\eta\kappa}\Big)^2\frac{\langle \hat{n}_{in}\rangle}{\eta},
\end{equation}
which is shown in Fig.~\ref{fig:F2}(a).

\textit{The optimal classical strategy.}---The negative effect of saturation on classical probe efficiency, as seen by the downward trend of the red lines in Fig.~\ref{fig:F2}(a), is visible for ${\kappa\ll1}$ suggesting sample saturation has a measurable impact on schemes probed far below saturation. For a given sample of known length and estimated absorption, we can maximise the FI over probe power. We find the optimal probe intensity $n_{opt}$ to be:
\begin{equation}\label{optN}
    n_{opt}=\frac{n_s}{2}\mathcal{W}[1+aL]\geq\frac{n_s}{2},
\end{equation}
with a corresponding sample transmission of  ${\eta_{opt}=\mathcal{W}[1+aL]^{-1}}$. Interestingly, the probe power resulting in the greatest precision is lower bounded by ${\kappa=1}$ implying classical strategies must be probed in the saturation regime to fully optimise performance. This result highlights an inherent trade-off between damage which can occur with saturation and desired precision that under-performing classical schemes will need to navigate. Such a compromise further motivates a move to the quantum regime which is capable of providing an absolute advantage~\cite{Casacio2021Quantum-enhancedMicroscopy}.

To highlight the benefits of classical power optimisation, we apply our method to the quantum-limited high-precision absorption spectroscopy of a dilute Cesium vapour cell, used to define the Boltzmann constant $k_z$ via Doppler broadening thermometry (DBT)~\cite{Truong2015AccurateConstant}. In DBT the characterisation of the line-width profile of a specific transition enables precise estimation of the Boltzmann constant $k_z$. DBT is also frequently used to accurately detect and monitor gasses~\cite{Bain2011RecoveryInvestigation} with high measurement precision being imperative to both applications of the technique. In Ref.~\cite{Truong2015AccurateConstant}, a shot-noise limited \SI{895}{\nano\meter} laser is used to probe a transition with characteristic saturation intensity ${n_s =}$~\SI{2.5}{\milli\watt\per\centi\meter^2} through a cell of length \SI{75}{mm}. A laser intensity of \SI{7}{\micro\watt\per\centi\meter^2} (${\kappa=0.005}$) is used resulting in a total transmission of ${\eta=17\%}$. We can therefore estimate the linear absorption coefficient to be ${a = }$~\SI{2.53}{\centi\meter^{-1}}. Using Eq.~\ref{optN}, we find the optimal probe intensity to be \SI{2.67}{\milli\watt\per\centi\meter^2}~(${\kappa = 2.14}$) with a resulting sample transmission of ${\eta= 47\%}$. Employing such a power would result in a \SI{24}{\decibel} improvement in the linear absorption estimate precision (total Fisher information) over the power used in~\cite{Truong2015AccurateConstant}. In the absence of saturation, a further \SI{6}{dB} increase in power would result in a \SI{6}{dB} increase in precision. Our model shows that once saturation is properly accounted for such an increase in probe power actually results in a \SI{-3}{dB} reduction in precision highlighting the importance of including saturation. Note, the only source of noise we consider here is laser shot-noise. Other sources of noise may scale unfavourably with power (e.g. temperature stability) and therefore may limit the practicality of witnessing such an increase in precision. This example demonstrates the potential gains in accounting for saturation when optimising measurements and highlights the potential for miscalculation when only considering linear absorption.

For a given intensity damage threshold, we can similarly maximise the FI over sample length. As a further example, we investigate the resonant absorption of Chlorophyll A Acetone solution, probed at \SI{661}{\nano\meter}~\cite{Lichtenthaler2001ChlorophyllsSpectroscopy}. Chlorophyll A density is routinely measured via absorption spectroscopy in a wide variety of settings~\cite{Aminot200ICESMer,Schalles2006OpticalConcentrations}. The transition absorption cross-section is ${\sigma=4\times10^{-17}}$~\si{cm^{2}} with lifetime ${\tau=}$~\SI{4}{\nano\second}~\cite{Correa2002, Correa2007ExcitedA}. Chlorophyll absorption measurements use a typical cuvette width of \SI{1}{\centi\meter} and aim to prepare sample densities that give an output transmission somewhere in the range of ${15\%-50\%}$~\cite{Lichtenthaler2001ChlorophyllsSpectroscopy}. Suppose we probe a sample with a commercially available high power laser at \SI{1}{\watt} (${\kappa=0.02}$) and measure a transmission of ${50\%}$. We can infer from this measurement that ${a\approx}$~\SI{0.7}{\centi\meter^{-1}}. The optimal sample length of such a measurement is found to be ${L=}$~\SI{2.9}{\centi\meter} which results in a \SI{3}{\decibel} improvement in precision over the standard \SI{1}{\centi\meter} cuvette width. This precision improvement maps directly onto the concentration estimate. As demonstrated above, these results provide simple but powerful optimisation strategies to help improve precision. We now derive a limit on the precision gained via any single-mode state and explore the ability of quantum states to saturate this bound.
\begin{figure*}
    \includegraphics[scale = 0.85]{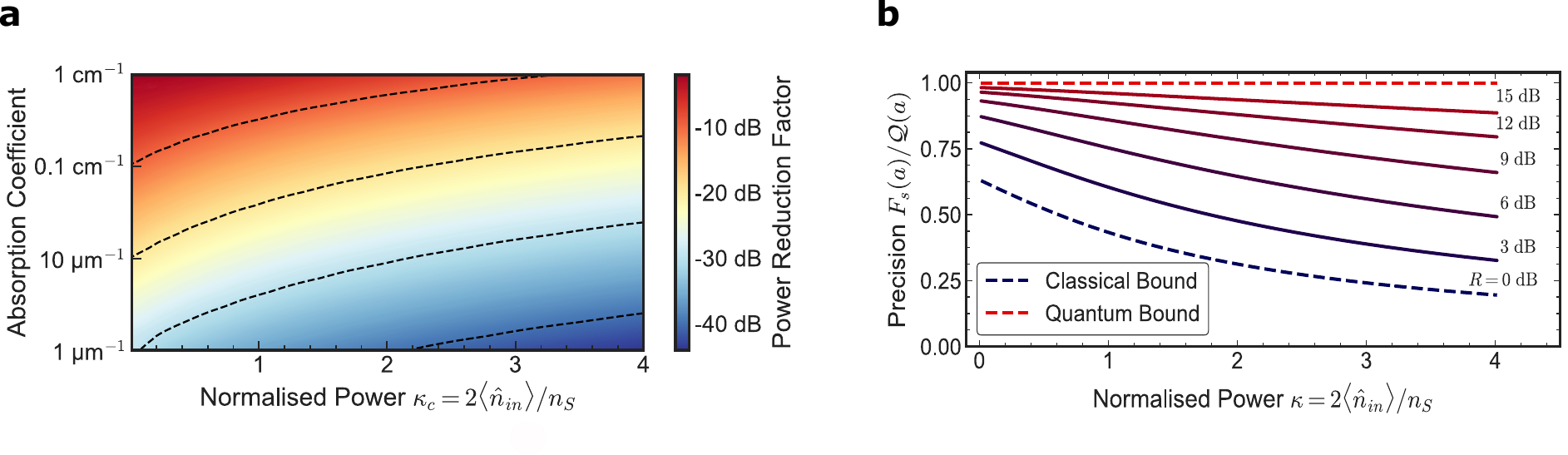}
    \caption{(a) The probe power reduction achievable by switching from a coherent probe of input intensity $\kappa_c$ to a Fock state probe without loss of precision. The sample has a fixed length $L=$~\SI{1}{\centi\meter} and is plotted for several absorption coefficient. (b) The squeezed state precision $F_s(a)$ normalised by the quantum limit $\mathcal{Q}(a)$ displaying the relative performance of several amplitude-squeezed states with squeezing factor $R$.}
    \label{fig:F3}
\end{figure*}

\textit{A bound on single-mode precision.}---Birchall \textit{et. al.} derived an upper bound on the QFI for single parameter estimation that results in a correlated linear phase and loss being applied to a single mode optical probe by Eq.~\ref{equ:CPLE}~\cite{Birchall2020QuantumLoss}, which reduces to the following when there is no phase shift applied by the sample:
\begin{equation}\label{equ:CPLBound}
    \mathcal{Q}(x)\leq\frac{\langle\hat{n}_{in}\rangle(\partial_{x}\eta)^2}{{\eta(1-\eta)}}
\end{equation}

The saturation model presented above is insensitive to phase information and depends only on the mean intensity of the input state and sample parameters. Therefore, we can apply this result to our model and note that an optimal quantum probe is one that necessarily saturates this bound. Using Eq.~\ref{equ:CPLBound} in conjunction with Eq.~\ref{equ:TRANS}, we derive a bound on the QFI gained on the linear absorption coefficient $a$, estimated via the transmission $\eta$:
\begin{equation}\label{equ:CPLE}
\begin{split}
    \mathcal{Q}(a) &= \Big(\frac{L\eta}{1+\eta\kappa}\Big)^2\frac{\langle \hat{n}_{in}\rangle}{\eta(1-\eta)}.
\end{split}
\end{equation}
Equation~\ref{equ:CPLE} defines the quantum limit on precision (shown in Fig.~\ref{fig:F2}(b)). We now seek to find quantum strategies that saturate this bound and the separation of such strategies from those reliant on classical resources only.

\textit{Finding an optimal probe.}---It is well-known that the Fock state-probe provides an optimal strategy in the linear absorption model~\cite{Holland1993InterferometricLimit,Wolf2019MotionalIons}. We expect this probe to remain optimal under saturation. To prove this hypothesis, we combine Eq.~\ref{equ:CFI1} and Eq.~\ref{equ:LLC} for a Fock state (${\Var(\hat{n}_{in})=0}$) which gives an expression for the FI achieved via Fock state probing ${ F_f(a)}$: \begin{equation}\label{equ:FFI}
    F_f(a)=\Big(\frac{L\eta}{1+\eta\kappa}\Big)^2\frac{\langle \hat{n}_{in}\rangle}{\eta(1-\eta)}.
\end{equation}
The FI for a Fock state is indeed equivalent to Eq.~\ref{equ:CPLE}, hence Fock state probing remains an optimal quantum strategy for maximising the precision of the absorption coefficient estimate in the presence of saturation. By setting ${\kappa=0}$, we recover well-known linear absorption estimation results with new insight gained for all ${\kappa>0}$. 

To benchmark the optimal quantum strategy against an equally bright classical strategy we define the quantum advantage ${\Lambda(a):=\mathcal{Q}(a)/F_c(a)}$ explicitly given by ${\Lambda(a)=\frac{1}{1-\eta}}$, which compares the maximum precision gained from quantum and classical states of equal brightness (Fig.~\ref{fig:F2}(c)). By performing a Taylor expansion around ${a=}$~\SI{0}{\centi\meter^{-1}} we find an expression for the quantum advantage gained under the saturated probing of weakly absorbing samples:
\begin{equation}\label{equ:QA}
    \Lambda(a) = \frac{1+\kappa}{aL},
\end{equation}
valid for ${\kappa\geq1}$ and ${a\leq\frac{1+\kappa}{L}}$.

\textit{The effect of saturation on precision.}---Saturation imposes a limit on the maximum precision a coherent, (Fig.~\ref{fig:F2}(a)), or Fock, (Fig.~\ref{fig:F2}(b)), state provides with increasing optical power. This is diametrical to linear loss which suggests probe brightness can always be increased to enhance precision~\cite{Allen2020ApproachingResources}. The knock-on effect is impaired efficiency due to diminishing returns in precision. Whilst the probe power is above saturation, the effective linear absorption is low which limits the information on the loss coefficient carried by each photon. Both the quantum and classical strategies suffer due to this effect, however, the coherent state performance is affected further due to greater optical noise in the input state. The combined effect is a quantum advantage that scales linearly with $\kappa$ in the saturation regime. The quantum advantage is strongest for weakly absorbing samples as observed in trace detection schemes~\cite{Harrison1969TheSpectroscopy,Christian1969MedicineSpectroscopy} or single-molecule direct absorption schemes~\cite{Chong2010}.

\textit{Probe brightness.}---The superior performance in combination with resilience to saturation allows for probing with a Fock state at much weaker probe powers without a compromise in performance. This enables a significant reduction in the energy flux incident upon the sample. In Fig.~\ref{fig:F3}(a) we quantify this reduction in required power for a given linear absorption coefficient and classical probe power $\kappa_c$. For a target absorption of ${a=}$~\SI{1}{\micro\meter^{-1}} probed at ${\kappa=1}$, employing the optimal quantum probe offers a reduction in probe brightness of \SI{-36}{\decibel} which has the potential to drastically reduce sample damage. We can account for the effect of imperfect detection by adding in a static loss pre-detector (see Supplementary Material B). Returning to our example of DBT and accounting for a detector efficiency of ${85\%}$, if we allow the classical strategy to optimise its power such that ${\kappa = 2.14}$ as previously calculated, we find a possible power reduction factor of \SI{-3}{\decibel} available by switching to the optimal quantum probe without a compromise in precision. State-of-the-art multiplexed single-photon sources are not bright enough to provide the required power to match performance and typically have repetition rates of the order ${\sim}$~\SI{e6}{\Hz}~\cite{Meyer-Scott2020}. Such a source would only begin to saturate samples with relaxation rates ${\sim}$~\si{\micro\second} while typical biophysical relaxation rates $\sim$~\si{\nano\second}~\cite{Berezin2010,Eisaman2011}. Although current single-photon sources lack the brightness to outperform classical probes, bright amplitude-squeezed states are readily available and capable of approaching the limit on quantum performance~\cite{Teich1989SqueezedLight}.
\begin{figure*}
    \includegraphics[scale = 0.95]{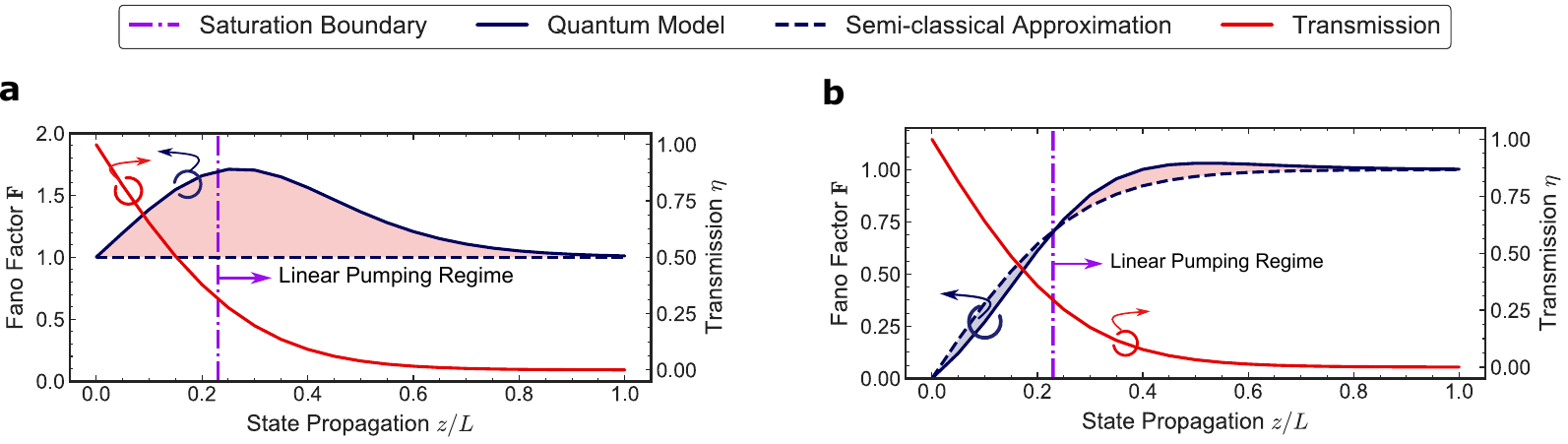}
    \caption{Coherent (a) and Fock (b) state evolution along a saturable sample. We plot the evolution of the probe state's Fano factor $\mathbf{F}$ (blue) and transmission $\eta$ (red) comparing the fully-quantum model (solid blue) to the semi-classical approximation (dashed blue). Shaded red represents additional noise and shaded blue represents suppressed noise. The dot-dashed purple line is the point along the channel at which the number of photons present in the probe state drops below $N_a$ (see Supplementary Material C).}
    \label{fig:F4}
\end{figure*}

\textit{Amplitude-squeezed state performance.}---The precision gained from such a bright amplitude-squeezed state, ${F_s(a)}$, is given by:
\begin{equation}
    F_s(a) = \Big(\frac{L\eta}{1+\eta\kappa}\Big)^2\frac{\langle \hat{n}_{in}\rangle}{\eta^{2}10^{-R/10}+\eta(1-\eta)}
\end{equation}
where $R$ is the input state's squeezing factor in \si{\decibel}, valid for bright amplitude-squeezing such that ${|\alpha|^2\gg R^{2}}$~\cite{Teich1989SqueezedLight}. Squeezed state performance is thus found to approach the quantum limit for infinite squeezing values,~${F_s(a)\xrightarrow[]{R\to\infty}\mathcal{Q}(a)}$. Fig.~\ref{fig:F3}(b) shows squeezed state performance normalised by $\mathcal{Q}(a)$. A state-of-the-art squeezed vacuum source with \SI{15}{\decibel} squeezing could be displaced to provide precision within $85\%$ of the quantum limit and is almost unaffected by saturation effects for ${\kappa\leq4}$~\cite{Vahlbruch2016}. We note that losses other than those caused by the target sample absorption, such as coupling losses or additional propagation losses, will dilute the squeezed statistics with vacuum noise, negatively affecting the state's performance. However, these results strongly indicate that advances can be made by employing squeezed states in the absence of bright Fock states.

\textit{Quantum loss channels.}---The semi-classical approximation of saturation is useful for providing analytical results. We now present a fully-quantum Hamiltonian model to asses the true effect of saturation on the higher order moments of the state's photon number. The model is presented in full in Supplementary Material C and has been solved using the open source software QuTip~\cite{Johansson2012QuTiP:Systems,Johansson2013QuTiPSystems}.

To enable full simulation, the sample is discritised along its length $z$. The probe state is propagated through each slice $dz$ containing $N_a$ absorbers and interacts with the probe state for a time ${\tau_{int}:=dz/c}$ where c is the speed of light inside the sample. Using open systems quantum modelling we account for each coherent absorber-field interaction, spontaneous emission and dephasing. The single-photon single-atom interaction rate is determined by the interaction time and field geometry relative to the absorption cross-section~\cite{Leuchs2012}. Quantum effects dominate the evolution if the absorber-field interaction rate exceeds all decohering effects~\cite{Hammerer2010,Stace2010TheoryVapor,Paris-Mandoki2017}. This type of coherent coupling is atypical for standard sensors and requires careful design. Here, we consider decoherence dominated propagation with a general dephasing time $T_2$ determined by the specific system under consideration. Using this model we can reproduce the classical dynamics of the ground and excited state populations. To compare the two models, we plot the evolution of the probe state's Fano factor -- the variance-to-mean ratio of the photon number probability distribution -- defined as ${\mathbf{F}:=\langle\hat{n}\rangle/\Var(\hat{n})}$, for classical (Fig.~\ref{fig:F4}(a)) and Fock (Fig.~\ref{fig:F4}(b)) input states.

As the state propagates along the sample, the photon number distribution evolves according to the effective loss applied to each component number state. In the fully quantum model, this effective loss is dependent on the number of photons in each basis state and is therefore different for each component of the distribution. By comparison, the semi-classical model assumes the effective loss is the same for all component number states and is dependent only on the mean photon number. The overall effect of accounting for the differential effective loss across the distribution is to add or suppress additional noise in comparison to the semi-classical approximation.

We find that the addition of noise to the Fock state is suppressed in comparison to the semi-classical model whilst the probe intensity remains above saturation. During a saturated interaction only $N_a$ photons may be absorbed per interaction time which suppresses the rate at which noise, through loss, is added to the state. ${F_f(a)}$ is therefore underestimated for output powers ${\kappa\geq1}$. For coherent state propagation, saturation increases the optical noise due to a differential effective absorption rate across the state's photon number distribution. This acts to stretch the Poissonian photon number statistics such that the probe becomes super-Poissonian. We therefore conclude that $F_{c}(a)$ presented under the semi-classical approximation is an upper bound on the true classical state performance. Consequently, the quantum advantage given by Eq.~\ref{equ:QA} is a lower bound on the true quantum advantage under saturated probing.

In alignment with the results presented by Kumar \textit{et al.} in ~\cite{Kumar}, the semi-classical approximation for classical sensing experiments is sufficient when the mean photon number of the state remains one standard deviation above or below the saturation intensity during the analyte-probe interaction. This ensures that the effective absorption strength across the states poissonian photon number statistics is well approximated by its mean value.

In conclusion, we have analysed and compared the effect of saturation on probe performance in absorption spectroscopy. These results are of most importance to saturation-limited classical measurements demanding greater precision or efficiency such as those often performed in high-precision absorption spectroscopy experiments~\cite{Truong2015AccurateConstant,Bain2011RecoveryInvestigation}. We have proven Fock states are optimal for mitigating the limiting effects of saturation and, through performance comparison, have shown classical schemes probing weak absorptions stand to gain the most from using a quantum probe~\cite{Carpentier1987,Bopp1997,NathanHenderson2007}. For schemes that do not have access to quantum resources we present a classical sample-probe optimisation strategy to help improve performance with minimal experimental adaptation. Further, we have shown that in the absence of bright Fock states, state-of-the-art squeezed states provide an effective alternative to overcoming saturation with today's technology. Whilst the analysis here finds saturation to be a detrimental effect on standard absorption spectroscopy, there are a number of more advanced techniques, such as stimulated depletion microscopy~\cite{Chong2010} or saturated structured-illumination microscopy~\cite{Ingerman2019SignalMicroscopy}, where saturation is used as a tool to enhance the information gained on a sample. Adapting the analysis here to optimise the performance of these strategies and to investigate any potential quantum advantages may lead to further enhancement of these optical sensors. Further work also includes experimental confirmation of these results. 

The authors would like to thank helpful and informative discussions with A. S. Clark, H. Gersen, and D. P. S. McCutcheon. We thank M. W. Mitchell for highlighting work related to this manuscript. This work was supported by the Centre for Nanoscience and Quantum Information (NSQI), EPSRC UK Quantum Technology Hub QUANTIC EP/M01326X/1. J.B. acknowledges support from EPSRC Quantum Engineering Centre for Doctoral Training EP/LO15730/1. S. W. acknowledges support from EPSRC grants EP/M024385/1 and EP/R024170/1. J.W.S acknowledges support from the Leverhulme Trust ECF-2018-276 and UKRI Future Leaders Fellowship MR/T041773/1. J.C.F.M. acknowledges support from an EPSRC Quantum Technology Fellowship EP/M024385/1 and an ERC starting grant ERC-2018-STG 803665. EJA acknowledges support from EPSRC doctoral prize EP/R513179/1.

\balance
\bibliographystyle{ieeetr}

\newpage

\onecolumngrid
\appendix

\section{Supplementary Material A: Classical Transmission}\label{sup:1}

We expand on the Beer-Lambert law to account for nonlinear absorption that occurs across a medium in the high power, low saturation intensity, regime. We aim to solve for the intensity of a single mode optical field, ${N(z)}$, as a function of position $z$ along a medium comprised of homogeneous two-level absorbers. The incident photons are resonant with the transition to remove frequency dependence and we consider the medium to be isolated from the environment, exhibiting no internal interactions or temperature dependent effects on population. Each particle has a given interaction cross section $\sigma$, the probability a particle will absorb a photon, and a transition relaxation time $\tau$. The total number of particles, $n_t$, is conserved with $n_0$ and $n_1$ accounting for the number of particles in the ground and excited state respectively. It follows that:

\begin{equation}\label{equ:Conservation}
    n_t=n_0+n_1.
\end{equation}

The rate equations for the populations of the ground and excited states at a given position z, accounting for absorption, stimulated emission, and spontaneous emission are:
\begin{equation}\label{equ:CDS}
    \begin{split}
        \frac{dn_0}{dt}=-\sigma N(z)n_0 + \sigma N(z)n_1 +\tau^{-1}n_1,\\
        \frac{dn_1}{dt}=\sigma N(z)n_0 - \sigma N(z)n_1 -\tau^{-1}n_1.
    \end{split}
\end{equation}
As the flux of photons passes through a slab of thickness $dz$, we make the assumptions that the relaxation time of the transition is long enough such that the populations of $n_1$ and $n_0$ can be considered constant. Under this steady-state assumption, ${\frac{dn_1}{dt}=0}$ and ${\frac{dn_0}{dt}=0}$. This allows us to solve for $n_1$ and $n_0$ using Eq.~\ref{equ:Conservation} and either one of \ref{equ:CDS}:
\begin{equation}\label{equ:CPE}
    \begin{split}
        n_0=\frac{n_t(\sigma N(z) + \tau^{-1})}{2\sigma N(z) + \tau^{-1}},
        \\
        n_1=\frac{n_t\sigma N(z)}{2\sigma N(z) + \tau^{-1}}.
    \end{split}
\end{equation}
As the state propagates along $dz$ the change in intensity is given by the net change in population:
\begin{equation}
    \frac{dN(z)}{dz}=\sigma N(z)(n_1-n_0).
\end{equation}
Here we assume spontaneous emission coupling back into the mode is negligible, as is the case for free space propagation. Using the solutions for $n_1$ and $n_0$ gives a differential in ${N(z)}$:
\begin{equation}
    \frac{dN(z)}{dz}=-\frac{\tau^{-1}n_t\sigma N(z)}{\tau^{-1}+2\sigma N(z)}.
\end{equation}

Let ${a:=n_t\sigma}$, the standard definition of the linear absorption coefficient, and ${n_s:= \frac{1}{\sigma\tau}}$ be the saturation intensity. Recast with these substitutions:
\begin{equation}
    \frac{dN(z)}{dz}=-\frac{aN(z)}{1+ 2N(z)/n_s}.
\end{equation}
The differential reduces to the Beer-Lambert law in the case of ${N(z)<<n_s/2}$. When this inequality is violated, the Beer-Lambert law is no longer an accurate model. We can now solve for ${N(z)}$ via separation of variables followed by integration from 0 to z. Let ${N(0):=N_0}$,
\begin{equation}\label{sequ:diff}
    e^{\frac{2N(z)}{n_s}
    }\frac{2N(z)}{n_s} = e^{ln(\frac{2N_0}{n_s} )+\frac{2}{n_s}(N_0 - az)}
\end{equation}
Eq.~\ref{sequ:diff} has been rearrange into a form with which we can use the Lambert-W function, defined by:
\begin{equation}
    W\big(ze^z) = z.
\end{equation}
Furthermore, using a special case of the Lambert-W function, the Wright Omega function, defined by ${W(e^z)=\mathcal{W}(z)}$, gives the final solution:
\begin{equation}
    N(z) = \frac{n_s}{2}\mathcal{W}\Big(ln\Big(\frac{2N_0}{n_s}\Big)+\frac{2N_0}{n_s} - a z\Big)\Big).
\end{equation}

The intensity through a sample of length L with linear absorption coefficient $a$, saturation intensity $n_s$ and input probe intensity ${N_0\equiv\langle \hat{n}\rangle_{in}}$ is therefore given by:
\begin{equation}
    \eta=\frac{n_s}{2\langle \hat{n}\rangle_{in}}\mathcal{W}\Big(ln\Big(\frac{2\langle \hat{n}\rangle_{in}}{n_s}\Big)+\frac{2\langle \hat{n}\rangle_{in}}{n_s} - a L\Big).
\end{equation}

\begin{figure}[!h]
\centering

    \includegraphics[scale = 0.4]{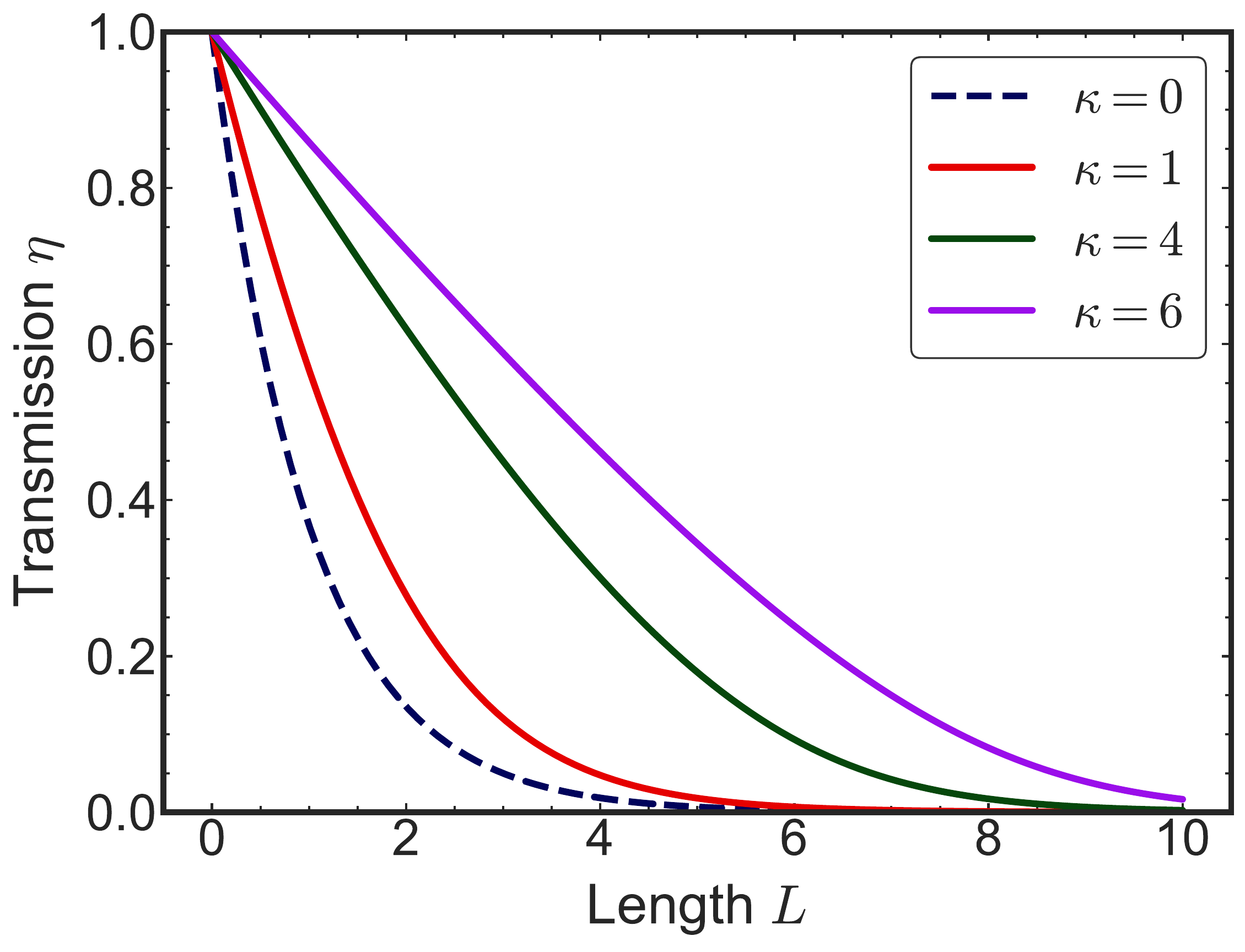}

    \caption{Sample transmission as a function of length for several probe intensities, $\kappa$. $\kappa =0$ recovers the Beer-Lambert linear absorption model, $\kappa = 1$ defines the edge of the saturation regime and all $\kappa>1$ is in the saturation regime.}
    
    \label{fig:SA_STR_KAPPA}
    
\end{figure}

We introduce ${\kappa=2\langle \hat{n}\rangle_{in}/n_s}$ as a dimensionless quantity:
\begin{equation}
    \eta(\kappa)=\frac{1}{\kappa}\mathcal{W}(ln(\kappa)+\kappa-aL).
\end{equation}
Fig.~\ref{fig:SA_STR_KAPPA} plots how transmission is effected by probing above and below saturation. As we increase $\kappa$, the transmission function shifts from an exponential, as determined by the Beer-Lambert law, to a linear fall-off conducive of saturated absorption.

\section{Supplementary Material B: Fisher Information}\label{sup:2}

To calculate the Fisher information (FI) gained from a probe state ${\ket{\psi}}$ on the unknown parameter $a$, we invoke the Cramér-Rao bound (CRB). Transmission is the optimal unbiased estimator for the linear absorption coefficient such that the FI on $\eta$ is given by:
\begin{equation}
    F(\eta) = \frac{1}{\Var(\eta)}.
\end{equation}
Given that ${\eta:=\langle \hat{n}\rangle/\langle \hat{n}_{in}\rangle}$ is a continuous linear differentiable function of output intensity ${\langle \hat{n}\rangle}$, we can relate the variance in the transmission to the variance of the output photon number via error propagation:
\begin{equation}
\begin{split}
    \Var(\eta) &= \left(\frac{\partial \eta}{\partial\hat{n}}\right)^2 \Var(\hat{n})\\
    &= \frac{\Var(\hat{n})}{\langle \hat{n}_{in}\rangle^2}.
\end{split}
\end{equation}
The FI is therefore given in terms of the output photon number variance of the output state:
\begin{equation}
    F(\eta) = \frac{\langle \hat{n}\rangle_{in}^2}{Var(\hat{n})}.
\end{equation}
The FI on $\eta$ can be related to the FI on $a$ via the formula:
\begin{equation}
\begin{split}\label{sequ:fish} 
    F(a)&=\Big(\frac{\partial\eta}{\partial a}\Big)^2 F(\eta)\\
    &= \Big(\frac{\partial\eta}{\partial a}\Big)^2 \frac{\langle \hat{n}\rangle_{in}^2}{Var(\hat{n})}.
\end{split}
\end{equation}

Using the following formula for the derivative of the Wright Omega function:
\begin{equation}
    \frac{d\mathcal{W}(x)}{dx}=\frac{\mathcal{W}(x)}{1+\mathcal{W}(x)}.
\end{equation}
We can calculate the derivative of $\eta$ via a substitution ${u=ln(\kappa)+\kappa -aL}$:
\begin{equation}\label{sequ:properror}
    \begin{split}
        \frac{\partial\eta}{\partial a} &= -\frac{L}{\kappa}\frac{\partial\mathcal{W}}{\partial u}\\ &=-\frac{L\eta}{1+\eta\kappa}.
    \end{split}
\end{equation}
Combining Eq.~\ref{sequ:fish} with Eq.~\ref{sequ:properror} gives the final expression for the FI gained by a state ${\ket{\psi}}$ on $a$ as a function of output state photon number variance:
\begin{equation}
    F(a)=\Big(\frac{L\eta}{1+\eta\kappa}\Big)^2\frac{\langle \hat{n}\rangle_{in}^2}{\Var(\hat{n})}.
\end{equation}

We can account for the effect of imperfect detection by adding an additional loss to the transmission pre-detection. We model this via a beam splitter with reflectivity ${1-\gamma}$, where $\gamma$ is the quoted detector efficiency. Under this model, the measured transmission is defined as ${\eta_m:=\eta\gamma}$. The FI on the measured transmission is again given by the CRB:
\begin{equation}
    F(\eta_m) = \frac{1}{\Var(\eta_m)}.
\end{equation}
Using the formula $\Var(aX)=a^2\Var(X)$ gives:
\begin{equation}
    F(\eta_m) = \frac{1}{\gamma^2\Var(\eta)}.
\end{equation}

Following the previous method, the FI on the measured transmission can be related to the FI on the linear coefficient via:
\begin{equation}
    F(a)=\Big(\frac{L\eta\gamma}{\gamma+\eta\kappa}\Big)^2\frac{\langle \hat{n}\rangle_{in}^2}{\Var(\hat{n})}.
\end{equation}

\section{Supplementary Material C: Quantum Loss Channel}\label{sup:3}

We present a quantum model of a saturated loss channel. To allow for a fully quantum simulation the loss channel is discretize along it's length z into slices each consisting of $N_a$ absorbers. The optical probe state is then propagated through each slice to show how quantum noise is propagated along the sample. Each slice interacts for a time ${\tau_{int}=dz/c}$ where $c$ is the speed of propagation. The model used is closely related to the Dicke model~\cite{Garraway2011} of an open quantum system under a Markovian evolution~\cite{Manzano2020AEquation}. We apply this model to calculate the evolution of the probe field accounting for both coherent and decoherent processes. Sensors that have not been specifically designed to operate coherently will be dominated by decoherent processes that result from spontaneous emission, variational atomic trajectories and inhomogenous broadening across the ensemble. Since this is by far the most common type of dynamic seen in direct absorption schemes we focus our discussion on this. Specifically we account for decoherence due to the spontaneous emission of each independent absorber at a rate ${\Gamma_{sp}:=1/T_1}$ and due to the dephasing at a rate ${\Gamma_{dp}:=1/T_2}$. Here, ${T_1}$~\&~${T_2}$ are the characteristic energy and phase relaxation time scales of the specific system under consideration~\cite{MITOpenCourseWare2006ChapterMatter}. Our model follows the dipole, rotating wave and two-level system approximations. By assuming a homogeneous sample density and probe intensity, we can approximate each absorber to be equally coupled to the field which is considered to be resonant with the absorber transition frequency $\omega$. In the second-quantised form, the governing system Hamiltonian is found to be~\cite{Garraway2011}:

\begin{equation}
    \hat{H}_{S}=\hbar\omega\hat{a}^\dagger\hat{a} + \sum_{k=1}^{N_a}\Big\{
\hbar\omega\hat{\sigma}_k^\dagger\hat{\sigma}_k+g(\hat{a}\hat{\sigma}^\dagger_k+\hat{a}^\dagger\hat{\sigma}_k)\Big\}
\end{equation}

where ${\hat{\sigma}^\dagger_k}$ is the excitation operator for the ${k^{th}}$ absorber which has the form ${\hat{\sigma}_k = \ket{g}\bra{e}}$ (these operators follow standard spin-1/2 algebra). ${\hat{a}}$ is the quantum oscillator probe field annihilation operator. The first term propagates the radiation field. The second term propagates the $N_a$ two-level systems. The third and fourth terms account for absorption and its reverse process, stimulated emission, at a rate determined by the single absorber-field interaction strength $g$. For monochromatic illumination of a two-level system under the dipole approximation, the interaction strength $g$ is given by:

\begin{equation}
g = \frac{\mu_{12}E}{\hbar}
\end{equation}

where ${\mu_{12}}$ is the dipole transition moment and $E$ is the energy per photon within the interaction volume. There is some debate over the definition of $E$ in free space, however, it is common practice to use the cavity based interaction definition where the volume $V$ is now given by the interaction volume defined by the probe optical mode cross-section and duration. As such, $E$ is given by~\cite{Rempe1993AtomsSpace}:

\begin{equation}
    E = \sqrt{\frac{\hbar\omega}{2\epsilon_0V}}.
\end{equation}

The free space transition cross-section and spontaneous emission rates can be recast in terms of the dipole transition moment~\cite{Hilborn1982That}:

\begin{equation}
\sigma = \frac{\pi\omega\mu_{12}^2}{3\epsilon_0\hbar c},
\end{equation}

\begin{equation}
    \Gamma_{sp} = \frac{\omega^3\mu_{12}^2}{3\epsilon_0\pi\hbar c^3}.
\end{equation}

We note that deriving the dephasing rate for a given system is much more complex and so is typically measured experimentally for a given setup and is often found to be orders of magnitude greater than $g$~\cite{Anastopoulos2000}.
Following reference~\cite{Leuchs2012}, we express the optical mode area ${A=\pi r_{focus}^2}$ in multiples of the
wavelength ${r_{focus}:=\beta\lambda}$. We normalise the interaction time ${\tau_{int}}$ by the absorber lifetime ${\Gamma_{sp}}$ such that ${\tau_{int}:=\alpha/\Gamma_{sp}}$.
Using the following anzats for the interaction volume:

\begin{equation}
    V = \pi\beta^2\lambda^2\cdot c \cdot \tau_{int}
\end{equation}

we can express the interaction rate as:

\begin{equation}
    g = \sqrt{\frac{3}{2\alpha}}\frac{3}{\pi^2\beta}.
\end{equation}

Crucially, in the free space model, the coupling strength is dependent on the interaction time through the parameter $\alpha$. Note that in this model we do not include the saturation intensity $n_s$ as this only arises in the semi-classical model as a direct result of the constraint that only one photon may be absorbed per cross-section per transition lifetime. Here, this is already built into the Hamiltonian. As mentioned above, the strategy will be to use this model to propagate the probe through each $i$-th sample slice of width ${\delta z}$. The input density matrix of the ${0^{th}}$ slice is given by:

\begin{equation}
    \rho_{0}(0) = \ket{\psi_0(0)}\bra{\psi_0(0)} \otimes (\ket{0^k}\bra{0^k})
\end{equation}

where the optical input state ${\ket{\psi_0(0)}\in\{\ket{\alpha}, \ket{N}\}}$ is either a Fock state or a coherent state with mean photon number ${\langle n\rangle_{in}}$. The absorbers are assumed to be in the ground state. The evolution of the system density matrix ${\rho_i(t)}$, where the interaction with the environment has been traced out, is governed by the following master equation. Note we have switched to natural units in which ${\hbar\equiv1}$ and ${c\equiv1}$, we also set ${\omega = 1}$ such that ${\lambda = 2\pi}$. This is for convenience and does not effect the physics:

\begin{equation}\label{equ:LB}
    \frac{\partial\rho_i(t)}{\partial t} = -i\Big[\hat{a}^\dagger\hat{a} + \sum_{k=1}^{N_a}\Big\{
\hat{\sigma}_k^\dagger\hat{\sigma}_k+g( \hat{a}\hat{\sigma}^\dagger_k+\hat{a}^\dagger\hat{\sigma}_k), \rho_i\Big\}\Big] + \Gamma_{sp}\sum_{k=1}^{N_a} \big(\hat{\sigma}_k\rho_i\hat{\sigma}^\dagger_k - \frac{1}{2}\{\hat{\sigma}^\dagger_k\hat{\sigma}_k, \rho_i\}\big) + \Gamma_{dp}\sum_{k=1}^{N_a} \big(\hat{\sigma_z}_k\rho_i\hat{\sigma_z}^\dagger_k - \frac{1}{2}\{\hat{\sigma_z}^\dagger_k\hat{\sigma_z}_k, \rho_i\}\big).
\end{equation}

Here, ${\hat{\sigma_z}_k}$ is the Pauli-z dephasing operator for the ${k^{th}}$ atom. We propagate the system for a time ${\tau_{int}}$ and perform a partial trace over the absorbers Hilbert space to extract the output optical state from the $i$-th slice:

\begin{equation}
    \ket{\psi_i(\tau_{int})}\bra{\psi_i(\tau_{int})} = tr_{N_a}(\rho_i(\tau_{int})).
\end{equation}

This state is then coupled forward as the input state of the ${i+1^{th}}$ slice and the algorithm is thus repeated for the full length of the sample:

\begin{equation}
    \rho_{i+1}(0) = \ket{\psi_i(\tau_{int})}\bra{\psi_i(\tau_{int})} \otimes (\ket{0^k}\bra{0^k}).
\end{equation}

At each step, we calculate the states photon number variance and compare it directly with that predicted by a linear loss channel. For the purpose of this investigation we set ${\tau_{int}=1}$ and assume ${\alpha=0.5}$ such that the interaction time is half the transition relaxation time allowing saturation to take effect during the interaction. The coupling constant can no longer be considered the same for each absorber close to diffraction limited focusing due to the states polarisation becoming non uniform~\cite{Kimble2001}. We therefore consider a maximal focusing of ${\beta=10}$. This gives an interaction rate ${g=0.1}$ in natural units. We set the dephasing rate to be ${\Gamma_{dp}=2}$ such that decoherence dominates the evolution with ${\Gamma_{dp}\gg g, \Gamma_{sp}}$. The model can be used to recreate the famous optical Bloch equations under a classical probe approximation which describe the evolution of an ensemble of absorbers coupled to a classical coherent field~\cite{Liu2017StrongLimit}. Here, we will use the model to probe the intensity profile along a sample consisting of $20$ slices. To enable full simulation of the Hilbert space ${N_a=4}$ for each slice and we probe with an optical state consisting of ${\langle \hat{n} \rangle_{in}=12}$ input photons. Despite the simplicity of this model, we can use it to lend insight into how probing a loss channel above saturation effects quantum noise in the probe state.

\end{document}